\def\year{2022}\relax
\documentclass[letterpaper]{article} 
\usepackage{aaai22}  
\usepackage{times}  
\usepackage{helvet}  
\usepackage{courier}  
\usepackage[hyphens]{url}  
\usepackage{graphicx} 
\usepackage{natbib}  
\usepackage{caption} 
\DeclareCaptionStyle{ruled}{labelfont=normalfont,labelsep=colon,strut=off} 
\usepackage{algorithm}
\usepackage{algorithmic}
\usepackage{newfloat}
\usepackage{listings}
\usepackage{expl3}
\usepackage{xspace}

\makeatletter
\DeclareRobustCommand\onedot{\futurelet\@let@token\@onedot}
\def\@onedot{\ifx\@let@token.\else.\null\fi\xspace}

\def\eg{\emph{e.g}\onedot} 
\def\ie{\emph{i.e}\onedot} 
 
\def\etc{\emph{etc}\onedot} \def\vs{\emph{vs}\onedot}

\makeatother
\usepackage[switch]{lineno}
\usepackage{subfigure}
\usepackage{amsmath}
\usepackage{colortbl}
\usepackage{amssymb}
\usepackage{multirow}
\usepackage{arydshln}
\usepackage{bm}
\usepackage{booktabs}
\newcommand{\Th}[1]{\textsc{#1}}
\newcommand{\mr}[2]{\multirow{#1}{*}{#2}}
\newcommand{\mc}[2]{\multicolumn{#1}{c}{#2}}
\newcommand{\tb}[1]{\textbf{#1}}
\newcommand{\tabincell}[2]{\begin{tabular}{@{}#1@{}}#2\end{tabular}}

\floatstyle{ruled}
\newfloat{listing}{tb}{lst}{}
\floatname{listing}{Listing}

\title{Learning Token-based Representation for Image Retrieval}
\author {
	Hui Wu\textsuperscript{\rm 1},~
	{Min Wang\textsuperscript{\rm 2}\footnotemark[1],~
	Wengang Zhou\textsuperscript{\rm 1,2}}\thanks{Corresponding Author},~
	Yang Hu\textsuperscript{\rm 1},~
	Houqiang Li\textsuperscript{\rm 1,2} \\
}
\affiliations {
	\textsuperscript{\rm 1} CAS Key Laboratory of GIPAS, University of Science and Technology of China \\
	\textsuperscript{\rm 2} Institute of Artificial Intelligence, Hefei Comprehensive National Science Center \\
	wh241300@mail.ustc.edu.cn, wangmin@iai.ustc.edu.cn, \{zhwg, eeyhu, lihq\}@ustc.edu.cn
}

\begin{document}
\maketitle
\begin{abstract}
	In image retrieval, deep local features learned in a data-driven manner have been demonstrated effective to improve retrieval performance. 
	To realize efficient retrieval on large image database, some approaches quantize deep local features with a large codebook and match images with aggregated match kernel.
	However, the complexity of these approaches is non-trivial with large memory footprint, which limits their capability to jointly perform feature learning and aggregation.
	To generate compact global representations while maintaining regional
	 matching capability, we propose a unified framework to jointly learn local feature representation and aggregation.
	In our framework, we first extract deep local features using CNNs. 
	Then, we design a tokenizer module to aggregate them into a few visual tokens, each corresponding to a specific visual pattern.
	This helps to remove background noise, and capture more discriminative regions in the image. 
	Next, a refinement block is introduced to enhance the visual tokens with self-attention and cross-attention. Finally, different visual tokens are concatenated to generate a compact global representation. 
	The whole framework is trained end-to-end with image-level labels. 
	Extensive experiments are conducted to evaluate our approach, which outperforms the state-of-the-art methods on the Revisited Oxford and Paris datasets. Our code is available at \url{https://github.com/MCC-WH/Token}.
\end{abstract}

\begin{figure}[t]
	\begin{center}
		\includegraphics[width=0.99\linewidth]{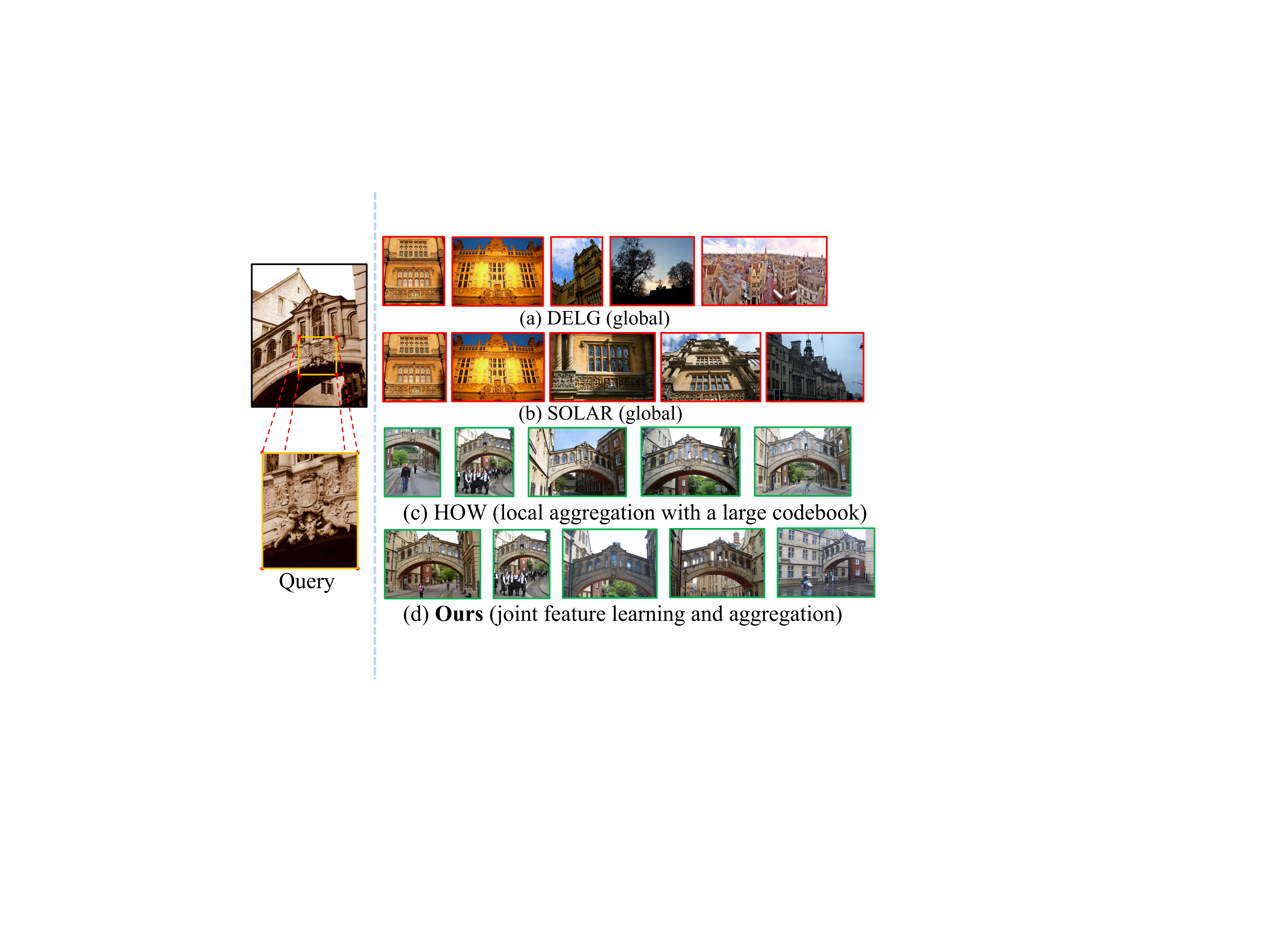}
	\end{center}
	\caption{Top-5 retrieval results of different methods, including DELG~\cite{cao2020unifying}, SOLAR~\cite{ng2020solar}, HOW~\cite{tolias2020learning} and ours. Query image is on the left (black outline) with a target object (orange box), and the right are the top-ranking images for the query. Our approach achieves similar results as HOW, which use large visual codebook to aggregate local features, with \textit{lower memory and latency}. Green solid outline: positive images for the query; red solid outline: negative results.}
	\label{fig:example}
\end{figure}

\section{Introduction} \label{sec:introduction}
Given a large image corpus, image retrieval aims to efficiently find target images similar to a given query.
It is challenging due to various situations observed in large-scale dataset, \eg, occlusions, background clutter, and dramatic viewpoint changes. 
In this task, image representation, which describes the content of images to measure their similarities, plays a crucial role. With the introduction of deep learning into computer vision, significant progress~\cite{babenko2014neural,gordo2017end,cao2020unifying,aaai1,noh2017large}  has been witnessed in learning image representation for image retrieval in a data-driven paradigm. 
Generally, there are two main types of representation for image retrieval. 
One is global feature, which maps an image to a compact vector, while the other is local feature, where an image is described with hundreds of short vectors. 

In global feature based image retrieval~\cite{radenovic2018fine,Babenko_2015_ICCV}, although the representation is compact, it usually lacks capability to retrieve target images with only partial match.  
As shown in Fig.~\ref{fig:example} (a) and (b), when the query image occupies only a small region in the target images, global features tend to return false positive examples, which are somewhat similar but do not indicate the same instance as the query image. 

Recently, many studies have demonstrated the effectiveness of combining deep local features~\cite{tolias2020learning,noh2017large,teichmann2019detect} with traditional ASMK~\cite{tolias2013aggregate} aggregation method in dealing with background clutter and occlusion. 
In those approaches, the framework usually consists of two stages: feature extraction and feature aggregation, where the former extracts discriminative local features, which are further aggregated by the latter for the efficient retrieval. 
However, they require offline clustering and coding procedures, which lead to a considerable complexity of the whole framework with a high memory footprint and long retrieval latency. 
Besides, it is difficult to jointly learn local features and aggregation due to the involvement of large visual codebook and hard assignment in quantization. 

Some existing works such as NetVLAD~\cite{arandjelovic2016netvlad} try to learn local features and aggregation simultaneously.
They aggregate the feature maps output by CNNs into compact global features with a learnable VLAD layer. Specifically, they discard the original features and adopt the sum of residual vectors of each visual word as the representation of an image. However, considering the large variation and diversity of content in different images, these visual words are too coarse-grained for the features of a particular image. This leads to insufficient discriminative capability of the residual vectors, which further hinders the performance of the aggregated image representation.

To address the above issues, we propose a unified framework to jointly learn and aggregate deep local features. 
We treat the feature map output by CNNs as original deep local features.
To obtain compact image representations while preserving the regional matching capability, we propose a tokenizer to adaptively divide the local features into groups with spatial attention. These local features are further aggregated to form the corresponding visual tokens. Intuitively, the attention mechanism ensures that each visual token corresponds to some visual pattern and these patterns are aligned across images. Furthermore, a refinement block is introduced to enhance the obtained visual tokens with self-attention and cross-attention. Finally, the updated attention maps are used to aggregate original local features for enhancing the existing visual tokens. The whole framework is trained end-to-end with only image-level labels.

Compared with the previous methods, there are two advantages in our approach. 
First, by expressing an image with a few visual tokens, each corresponding to some visual pattern, we implicitly achieve local pattern alignment with the aggregated global representation. As shown in Fig.~\ref{fig:example} (d), our approach performs well in the presence of background clutter and occlusion. 
Secondly, the global representation obtained by aggregation is compact with a small memory footprint.  
These facilitate effective and efficient semantic content matching between images. 
We conduct comprehensive experiments on the Revisited Oxford and Paris datasets, which are further mixed with one million distractors.
Ablation studies demonstrate the effectiveness of the tokenizer and the refinement block.
Our approach surpasses the state-of-the-art methods by a considerable margin. 

\begin{figure*}[t]
	\begin{center}
		\includegraphics[width=1.0\linewidth]{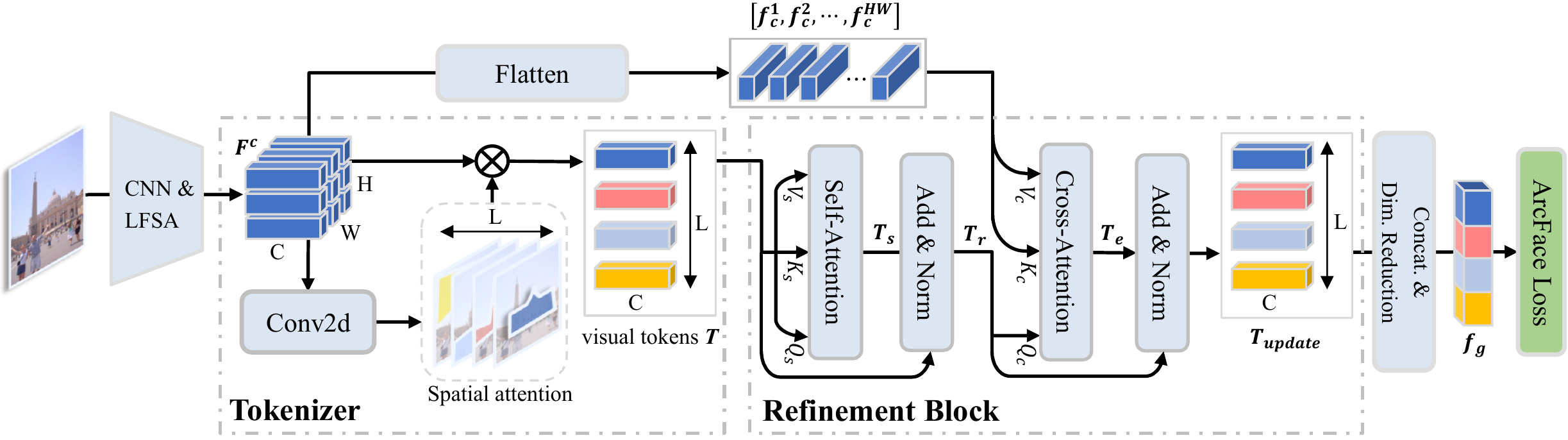}
	\end{center}
	\caption{An overview of our framework. 
		Given an image, we first use a CNN and a Local Feature Self-Attention (LFSA) module to extract local features $\bm{F}^c$. Then, they are tokenized into $L$ visual tokens with spatial attention. 
		Further, a refinement block is introduced to enhance the obtained visual tokens with self-attention and cross-attention.
		Finally, we concatenate all the visual tokens to form a compact global representation $\bm{f}_g$ and reduce its dimension.}
	\label{fig:framwork}
\end{figure*}

\section{Related Work}\label{sec:relation work}
In this section, we briefly review the related work including local feature and global feature based image retrieval.

\noindent\textbf{Local feature}.
Traditionally local features~\cite{lowe2004distinctive,bay2006surf,liu2015uniting} are extracted using hand-crafted detectors and descriptors. 
They are first organized in bag-of-words~\cite{sivic2003video,zhou2010spatial} and further enhanced by spatial validation~\cite{philbin2007object}, hamming embedding~\cite{hammingembeding} and query expansion~\cite{chum2007total}. 
Recently, tremendous advances~\cite{HardNet2017,tolias2020learning,AffNet2017,noh2017large,Tian_2019_CVPR,cao2020unifying,Wu_2021_ICCV} have been made to learn local features suitable for image retrieval in a data-driven manner. 
Among these approaches, the state-of-the-art approach is HOW~\cite{tolias2020learning}, which uses attention learning to distinguish deep local features with image-level annotations. During testing, it combines the obtained local features with the traditional ASMK~\cite{tolias2013aggregate} aggregation method. 
However, HOW cannot jointly learn feature representation and aggregation due to the very large codebook and the hard assignment during the quantization process. 
Moreover, its complexity is considerable with a high memory footprint. 
Our method uses a few visual tokens to effectively represent image. The feature representation and aggregation are jointly learned.

\noindent\textbf{Global feature}.
Compact global features reduce memory footprint and expedite the retrieval process.
They simplify image retrieval to a nearest neighbor search and extend the previous query expansion~\cite{chum2007total} to an efficient exploration of the entire nearest neighbor graph of the dataset by diffusion~\cite{aaai2}.
Before deep learning, they are mainly developed by aggregating hand-crafted local features, \eg, VLAD~\cite{jegou2011aggregating}, Fisher vectors~\cite{fisher_vector}, ASMK~\cite{tolias2013aggregate}. 
Recently, global features are obtained simply by performing the pooling operation on the feature map of CNNs. 
Many pooling methods have been explored, \eg, max-pooling (MAC)~\cite{tolias:hal-01842218}, sum-pooling (SPoC)~\cite{Babenko_2015_ICCV}, weighted-sum-pooling (CroW)~\cite{kalantidis2016cross}, regional-max-pooling (R-MAC)~\cite{tolias:hal-01842218}, generalized mean-pooling (GeM)~\cite{radenovic2018fine}, and so on. 
These networks are trained using ranking~\cite{radenovic2018fine,revaud2019learning} or classification losses~\cite{Deng_2019_CVPR}. 
Differently, our method tokenizes the feature map into several visual tokens, enhances the visual tokens using the refinement block, concatenates different visual tokens and performs dimension reduction. Through these steps, our method generates a compact global representation while maintaining the regional matching capability.

\section{Methodology} \label{sec:Methodology}
An overview of our framework is shown in Fig.~\ref{fig:framwork}. 
Given an image, we first obtain the original deep local features $\bm{F} \in \mathcal{R}^{C \times H \times W}$ through a CNN backbone. 
These local features are obtained with limited receptive fields covering part of the input image. 
Thus, we follow~\cite{ng2020solar} to apply the Local Feature Self-Attention (LFSA) operation on $\bm{F}$ to obtain context-aware local features $\bm{F}^c \in \mathcal{R}^{C \times H \times W} $. 
Next, we divide them into $L$ groups with spatial attention mechanism, and the local features of each group are aggregated to form a visual token $\bm{t} \in \mathcal{R}^{C}$. 
We denote the set of obtained visual tokens as $\bm{T} = \left[\bm{t}^{(1)},\bm{t}^{(2)},\cdots,\bm{t}^{(L)} \right] \in \mathcal{R}^{L \times C}$.
Furthermore, we introduce a refinement block to update the obtained visual tokens $\bm{T}$ based on the previous local features $\bm{F}^c$.
Finally, all the visual tokens are concatenated and we reduce its dimension to form the final global descriptor $\bm{f}_{g}$. ArcFace margin loss is used to train the whole network.
	
\subsection{Tokenizer}
To effectively cope with the challenging conditions observed in large datasets, such as noisy backgrounds, occlusions, \etc, image representation is expected to find patch-level matches between images.
A typical pipeline to tackle these challenges consists of local descriptor extraction, quantization with a large visual codebook created usually by $k$-means and descriptor aggregation into a single embedding.
However, due to the offline clustering and hard assignment of local features, it is difficult to optimize feature learning and aggregation simultaneously, which further limits the discriminative power of the image representation.

To alleviate this problem, we here use spatial attention to extract the desired visual tokens.
By training, the attention module can adaptively discover discriminative visual patterns.

For the set of local features $\bm{F}^c$, we generate $L$ attention maps $\bm{A}=\{\bm{a}^{(1)},\bm{a}^{(2)},\cdots,\bm{a}^{(L)}\}$, which are implemented by $L$ $1 \times 1$ convolutional layers. We denote the parameters of the convolution layers as $\mathbf{W} = \left[\mathbf{w}_1,\mathbf{w}_2,\cdots,\mathbf{w}_L\right] \in \mathcal{R}^{L\times C}$, and the attention maps are calculated as
\begin{eqnarray}\label{eq:atten}
	\begin{aligned}
		\bm{a}^{(i)}_{h,w} = \ \frac{\exp(\mathbf{w}_{i} \cdot \bm{F}^c_{h,w})}{\sum_{l=1}^{L}\exp(\mathbf{w}_{l} \cdot \bm{F}^c_{h,w})}.
	\end{aligned}
\end{eqnarray}
Then, the visual tokens $\bm{T} = \left[\bm{t}^{(1)},\bm{t}^{(2)},\cdots,\bm{t}^{(L)} \right]$ are computed as
\begin{eqnarray}\label{eq:token}
	\bm{t}^{(i)} = \frac{1}{\gamma(\bm{a}^{(i)})} \sum_{h \in H,w \in W} \bm{a}^{(i)}_{h,w}\bm{F}^c_{h,w},
\end{eqnarray}
where $\gamma(\bm{a}^{(i)}) = \sum_{h \in H,w \in W} \bm{a}^{(i)}_{h,w}$ and $\bm{t}^{(i)} \in \mathcal{R}^C$. 
 
\noindent\textbf{Relation to GMM}.
Tokenization aggregates local features into visual tokens, which helps to capture discriminative visual patterns, leading to a more general and robust image representation. 
The visual tokens represent several specific region patterns in an image, which share the similar rationale of learning a Gaussian Mixture Model (GMM) over the original local features of an image. GMM is a probabilistic model that assumes all the data points are generated from a mixture of a finite number of Gaussian distributions with unknown mean vectors and data variances. Formally,
\begin{eqnarray}\label{eq:gmm}
	\begin{aligned} 
		p(\bm{f}_i| z=j) = & \frac{1}{(2\pi \sigma_{j}^2)^\frac{C}{2}} \exp\left( - \frac{1}{2\sigma_j^2} \| \bm{f}_i - \mathbf{c}_j \|^2 \right), \\
		 p(\bm{f}_i) = & \sum_{j=1}^N p(z=j)p(\bm{f}_i| z=j).
	\end{aligned}
\end{eqnarray}
Here, $N$ is the number of Gaussian mixtures, $\bm{f}_i$ is the local feature of an image with dimension $C$. $z$ is the latent
cluster assignment variable, $\mathbf{c}_j$ and $\sigma_j$ correspond to the mean vector and variance of the $j$-th Gaussian distribution, respectively.
The Expectation-Maximization algorithm is commonly used to solve this problem. Iteratively, it estimates for each point a probability of being generated by each component of the model and updates the mean vectors:
\begin{eqnarray}\label{eq:gmm_update}
	\begin{aligned} 
		\mathbf{c}_j = & \frac{\sum_{i=1}^M p(z=j | \bm{f}_i)\bm{f}_i}{\sum_{i=1}^M p(z=j | \bm{f}_i)}, \ j=1,2,\cdots,N, \\
		p(z=j | \bm{f}_i) = & \frac{p(z=j) \times \exp\left( - \frac{1}{2\sigma_j^2} \| \bm{f}_i - \mathbf{c}_j \|^2 \right)}{\sum_{l=1}^N p(z=l) \times \exp\left( - \frac{1}{2\sigma_l^2} \| \bm{f}_i - \mathbf{c}_l \|^2 \right) },
	\end{aligned}
\end{eqnarray}
where $M$ is the total feature number. In GMM, $p(z=j | \bm{f}_i)$ represents the posterior probability of a local feature $\bm{f}_i \in \mathcal{R}^C$ being assigned to the $j$-th cluster.

In our approach, considering that $\mathbf{w}_{i} \cdot \bm{F}^c_{h,w} = \frac{1}{2}\| \bm{F}^c_{h,w}\|^2 + \frac{1}{2}\|\mathbf{w}_i\|^2 - \frac{1}{2}\| \bm{F}^c_{h,w} - \mathbf{w}_i\|^2$, Eq.~\eqref{eq:atten} can be reformulated as
\begin{eqnarray}\label{eq:re_atten}
	\bm{a}^{(i)}_{h,w} =\frac{\phi\left(\|\mathbf{w}_i\|^2\right) \times \exp\left( -\frac{1}{2} \| \bm{F}^c_{h,w} - \mathbf{w}_i \|^2\right)}{\sum_{l=1}^{L}\phi\left(\|\mathbf{w}_l\|^2\right) \times \exp\left( -\frac{1}{2}\| \bm{F}^c_{h,w} - \mathbf{w}_l\|^2\right)},
\end{eqnarray}
where $\phi\left(\|\mathbf{w}_i\|^2\right) = \exp(\frac{1}{2}\| \mathbf{w}_i\|^2) / \sum_{j=1}^{L}\exp(\frac{1}{2}\| \mathbf{w}_j\|^2)$.
We set $\sigma$ in Eq.~\eqref{eq:gmm} as 1 and $p(z=j)$ as $\phi\left(\|\mathbf{w}_i\|^2\right)$. $\bm{a}^{(i)}_{h,w}$ can be interpreted as a soft cluster assignment of local features $\bm{F}^c_{h,w}$ to $i$-th visual pattern, which is the same as the meaning of $p(z=j | \bm{f}_i)$. $\mathbf{w}_i$ is the mean vector corresponding to the $i$-th visual pattern, whose $L_2$ norm is proportional to the probability of occurrence of that visual pattern. Since $\mathbf{W}$ is the network parameter that learned from the image data distribution, it enables the alignment of the corresponding visual patterns across different images.
Further, as shown in Eq.~\eqref{eq:token}, the visual token $\bm{t}^{(i)}$ obtained by weighted aggregation are equivalent to the updated mean vectors $\mathbf{c}_j$ in GMM.

\subsection{Refinement Block}
The two major components of our refinement block are the relationship modeling and visual token enhancement. The former allows the propagation of information between tokens, which helps to produce more robust features, while the latter is utilized to relocate the visual patterns in the original local features and extract the corresponding features for enhancing the existing visual tokens $\bm{T}$.

\noindent \textbf{Relationship modeling}. 
During tokenization, different attention maps are used separately. It excludes any relative contribution of each visual token to the other tokens.
Thus, we employ the self-attention mechanism~\cite{tr} to model the relationship between different visual tokens $\bm{t}^{(i)}$, generating a set of relation-aware visual tokens $\bm{T}_{r}=\left[ \bm{t}_r^{(1)},\bm{t}_r^{(2)},\cdots,\bm{t}_r^{(L)}\right] \in \mathcal{R}^{L \times C}$.
Specifically, we first map visual tokens $\bm{T}$ to Queries ($\bm{Q}_{s} \in \mathcal{R}^{L \times C}$), 
Keys ($\bm{K}_{s} \in \mathcal{R}^{L \times C}$) and Values ($\bm{V}_{s} \in \mathcal{R}^{L \times C}$) with three $C \times C$-dimensional learnable matrices.
After that, the similarity $\bm{S} \in \mathcal{R}^{L \times L}$ between visual tokens is computed through
\begin{eqnarray}\label{equ:self_sim}
	\bm{S}(\bm{Q}_s,\bm{K}_s)=\text{\Th{Softmax}}(\frac{\bm{Q}_{s} \bm{K}_{s}^{T}}{\sqrt{C}})\ \in \mathcal{R}^{L \times L},
\end{eqnarray}
where the normalized similarity $\bm{S}_{i,j}$ models the correlation between different visual tokens $\bm{t}^{{i}}$ and $\bm{t}^{{j}}$.
To focus on multiple semantically related visual tokens simultaneously, we calculate similarity $\bm{S}$ with Multi-Head Attention (MHA). 

In MHA, different projection matrices for Queries, Keys, and Values are used for different heads, and these matrices map visual tokens to different subspaces. 
After that, $\bm{S}^{(i)}$ of each head, calculated by Eq.~\eqref{equ:self_sim}, is used to aggregate semantically related visual tokens. MHA then concatenates and fuses the outputs of different heads using the learnable projection $\bm{W}_M \in \mathcal{R}^{C \times C}$.
Formally,
\begin{eqnarray}\label{equ:MHA}
	\begin{aligned}
		\bm{T}_{s}^{(i)} =& \text{\Th{Dropout}}(\bm{S}^{(i)}\bm{V}_{s}^{(i)}),\ \mbox{for}\ i=1,2,\cdots,N, \\
		\bm{T}_{s} =& \text{\Th{Concat}}(\bm{T}_{s}^{(1)},\bm{T}_{s}^{(2)},\cdots,\bm{T}_{s}^{(N)})\bm{W}_M,
	\end{aligned}
\end{eqnarray}
where $N$ is head number and $\bm{T}_{s}^{(i)}$ is output of the $i$-th head.
Finally, $\bm{T}_{s}$ is normalized via Layer Normalization and added to original $\bm{T}$ to produce the relation aware visual tokens:
\begin{eqnarray}
	\label{equ:residue_connect_self}
	\bm{T}_r=\bm{T}+\text{\Th{LayerNorm}}(\bm{T}_{s}).
\end{eqnarray}

\noindent\textbf{Visual token enhancement}. 
To further enhance the existing visual tokens, we next extract features from $\bm{F}^c$ with the cross-attention mechanism.
As shown in Fig.~\ref{fig:framwork}, we first flatten $\bm{F}^c$ into a sequence $\left[ \bm{f}_c^{1},\bm{f}_c^{2},\cdots,\bm{f}_c^{HW} \right] \in \mathcal{R}^{HW \times C}$.
Then, with different fully-connected (FC) layers, $\bm{T}_r$ is mapped to Queries ($\bm{Q}_{c} \in \mathcal{R}^{L \times C}$) and $\bm{F}^c$ is mapped to Keys ($\bm{K}_{c} \in \mathcal{R}^{HW \times C}$) and Values ($\bm{V}_{c} \in \mathcal{R}^{HW \times C}$), respectively.
The similarity between the visual tokens $\bm{T}_r$ and the original local feature $\bm{F}^c$ is calculated as
\begin{eqnarray}
	\label{equ:cross_sim}
	\bm{S}(\bm{Q}_{c},\bm{K}_{c})=\text{\Th{Softmax}}(\frac{\bm{Q}_{c} \bm{K}_{c}^{T}}{\sqrt{C}})\ \in \mathcal{R}^{L \times HW}.
\end{eqnarray}
Here, the similarity $\bm{S}_{i,j}$ indicates the probability that the $j$-th local feature $f_c^{j}$ in $\bm{F}^c$ should be assigned to the $i$-th visual token, which is different from the meaning $\bm{S}$ in Eq.~\eqref{equ:self_sim}.
Then, the weighted sum of $\bm{F}^c$ and $\bm{S}$ is added to $\bm{T}_r$ to produce the updated visual tokens:
\begin{eqnarray}
	\label{equ:cross_add}
	\begin{aligned}
		\bm{T}_{c} =& \ \text{\Th{Dropout}}(\bm{S}\bm{V}_{c}), \\
		\bm{T}_{update}=& \ \bm{T}_r+\text{\Th{LayerNorm}}(\bm{T}_{e}).
	\end{aligned}
\end{eqnarray}
As in Eq.~\eqref{equ:MHA}, MHA is also used to calculate the similarity.

We stack $N$ refinement blocks to obtain more discriminative visual tokens. 
The refined visual tokens $\bm{T}_{update}\in\mathcal{R}^{L \times C}$ come from the output of the last block of our model.
We concatenate the different visual tokens $\bm{T}_{update}$ into a global descriptor and 
a fully-connected layer is adopted to reduce its dimension to $d$:
\begin{eqnarray}\label{equ:dim_reduce}
	\bm{f}_{g}=\text{\Th{Concat}}(\bm{t}_{update}^{(1)},\bm{t}_{update}^{(2)},\cdots,\bm{t}_{update}^{(L)})\bm{W}_g,
\end{eqnarray}
where $\bm{W}_g \in \mathcal{R}^{LC \times d}$ is the weight of the FC layer.

\subsection{Training Objectives}
Following DELG~\cite{cao2020unifying}, ArcFace margin loss~\cite{Deng_2019_CVPR} is adopted to train the whole model.
The ArcFace improves the normalization of the classifier weight vector $\hat{\bm{W}}$ and the interval of the additive angles $m$ so as to enhance the separability between classes and meantime enhance the compactness within class. 
It has shown excellent results for global descriptor learning by inducing smaller intra-class variance. Formally,
\begin{eqnarray}\label{ArcFace}
	\mathit{L} = -\log\left ( \frac{\exp\left ( \gamma \times {AF} \left (  \bm{\hat{w}}_{k}^{T}\hat{\bm{f}_g},1\right )\right )}{\sum_{n}^{}\exp\left ( \gamma \times {AF} \left (  \bm{\hat{w}}_{n}^{T}\hat{\bm{f}_g},y_{n}\right )\right )} \right ),
\end{eqnarray}
where $\bm{\hat{w_i}}$ refers to the $i$-th row of $\bm{\hat{W}}$ and $\bm{\hat{f_g}}$ is the $L_2$-normalized $\bm{f}_g$. $y$ is the one-hot label vector and $k$ is the ground-truth class index ($y_k=1$). $\gamma$ is a scale factor. 
$AF$ denotes the adjusted cosine similarity and it is calculated as:
\begin{eqnarray}\label{AF_COS}
	{AF}\left ( \mathit{s,c} \right ) =\ (1-c) \times s + c \times \cos\left ( \text{acos}\left ( s \right ) +m \right ),
\end{eqnarray}
where $s$ is the cosine similarity, $m$ is the ArcFace margin, and $c$ is a binarized value that denotes whether it is the ground-truth category.

\begin{table*}[h]
	\begin{center}
		\setlength{\extrarowheight}{0.5pt}
		\resizebox{\textwidth}{!}{
			\begin{tabular}{llcccccccc}
				\toprule
				\multicolumn{2}{c}{ \multirow{2}*{\Large \Th{Method}} }&  \multicolumn{4}{c}{\large \Th{Medium}} &\multicolumn{4}{c}{\large \Th{Hard}}\\
				\cline{3-6}  \cline{7-10} 
				\multicolumn{2}{c}{} & $\mathcal{R}$Oxf &
				$\mathcal{R}$Oxf+$\mathcal{R}$1M &
				$\mathcal{R}$Par &
				$\mathcal{R}$Par+$\mathcal{R}$1M &
				$\mathcal{R}$Oxf &
				$\mathcal{R}$Oxf+$\mathcal{R}$1M &
				$\mathcal{R}$Par &
				$\mathcal{R}$Par+$\mathcal{R}$1M  \\
				\toprule
				\multicolumn{10}{l}{\textsl{(A) Local feature aggregation} } \\ \hdashline
				\multicolumn{2}{l}{HesAff-rSIFT-ASMK$^\star$+SP}&60.60&46.80&61.40&42.30&36.70&26.90&35.00&16.80\\ 
				\multicolumn{2}{l}{DELF-ASMK$^\star$+SP(GLDv1-noisy)}&67.80&53.80&76.90&57.30&43.10&31.20&55.40&26.40\\ 	  			 		     
				\multicolumn{2}{l}{DELF-R-ASMK$^\star$+SP(GLDv1-noisy)}&76.00&64.00&80.20&59.70&52.40&38.10&58.60&58.60\\ 	
				\multicolumn{2}{l}{R50-HOW-ASMK$^\star$(SfM-120k)}& 79.40& 65.80 &81.60& 61.80  &56.90 & 38.90 & 62.40 & 33.70 \\ 
				\multicolumn{2}{l}{R101-HOW-VLAD$\dagger$(GLDv2-clean)}& 73.54& 60.38 &82.33& 62.56  &51.93 & 33.17 & 66.95 & 41.82 \\
				\multicolumn{2}{l}{R101-HOW-ASMK$^\star \dagger$(GLDv2-clean)}&\underline{80.42}& \underline{70.17} &\underline{85.43}& \underline{68.80}  &\underline{62.51} & \underline{45.36} &\underline{70.76}& \underline{45.39} \\	
				\toprule
				
				\multicolumn{10}{l}{\textsl{(B) Global features + Local feature re-ranking} } \\ \hdashline
				\multicolumn{2}{l}{R101-GeM+DSM}&65.30  &47.60 &77.40 &52.80 &39.20 &23.20 &56.20 &25.00 \\ 	
				\multicolumn{2}{l}{R50-DELG+SP(GLDv2-clean)}&78.30  &67.20 &85.70 &69.60 &57.90 &43.60 &71.00 &45.70 \\ 
				
				\multicolumn{2}{l}{R101-DELG+SP(GLDv2-clean)}&81.20  &69.10 &87.20 &71.50 &64.00 &47.50 &72.80 &48.70 \\
				\multicolumn{2}{l}{R101-DELG+SP$\dagger$(GLDv2-clean)}&\underline{81.78}  &\underline{70.12} &\underline{88.46} &\underline{76.04} &\underline{64.77} &\underline{49.36} &\underline{76.80} &\underline{53.69} \\ 
				\toprule
				
				\multicolumn{10}{l}{\textsl{(C) Global features} } \\ \hdashline
				\multicolumn{2}{l}{R101-R-MAC(NC-clean)}&60.90  &39.30 & 78.90&54.80 &32.40 &12.50 &59.40 &28.00 \\ 
				\multicolumn{2}{l}{R101-R-MAC$\dagger$(GLDv2-clean)}&75.14  &61.88 & 85.28&67.37 &53.77 &36.45 &71.28 &44.01 \\
				\multicolumn{2}{l}{R101-GeM-AP(GLDv1-noisy)}& 67.50  &47.50 &80.10 &52.50 &42.80 &23.20 &60.50 &25.10 \\ 
				\multicolumn{2}{l}{R101-SOLAR(GLDv1-noisy)}&69.90  & 53.50 &81.60 &59.20 & 47.90 &29.90 & 64.50 & 33.40 \\ 
				\multicolumn{2}{l}{R101-SOLAR$\dagger$(GLDV2-clean)}&\underline{79.65}  & 67.61 & \underline{88.63} &73.21 & 59.99 &41.14 & 75.26 & 50.98 \\ 
				\multicolumn{2}{l}{R50-DELG(GLDv2-clean)}&73.60  &60.60 &85.70 &68.60 &51.00 &32.70 &71.50 &44.40 \\ 
				\multicolumn{2}{l}{R50-DELG$\dagger$(GLDv2-clean)}&76.40  &64.52 &86.74 &70.71 &55.92 &38.60  &72.60 &47.39  \\
				\multicolumn{2}{l}{R101-DELG(GLDv2-clean)}&76.30  &63.70 &86.60 &70.60 &55.60 &37.50 &72.40 &46.90 \\ 
				\multicolumn{2}{l}{R101-DELG$\dagger$(GLDv2-clean)}&78.24  &\underline{68.36} &88.21 &\underline{75.83} &\underline{60.15} &\underline{44.41} &\underline{76.15} &\underline{52.40} \\ 
				\multicolumn{2}{l}{R101-NetVLAD$\dagger$(GLDv2-clean)}&73.91  &60.51 &86.81 &71.31 &56.45 &37.92 &73.61 &48.98 \\ 
				\toprule
				\multicolumn{2}{l}{\textbf{R50-Ours(GLDv2-clean)}} & 79.42 & 73.68 & 88.67 & 77.56 & 59.48 & 49.55 & 76.49 & 58.92\\
				\multicolumn{2}{l}{\textbf{R101-Ours$_{\mathbf{PQ8}}$(GLDv2-clean)}} & 82.02 & 75.17 & 89.16 & 78.83 & 65.90 & 50.46 & 78.07 & 59.72 \\
				\multicolumn{2}{l}{\textbf{R101-Ours$_{\mathbf{PQ1}}$(GLDv2-clean)}} & \textbf{82.30} & \textbf{75.68} & 89.33 & 79.76 & \textbf{66.62} & \textbf{51.65} & 78.55 & 61.54 \\
				\multicolumn{2}{l}{\textbf{R101-Ours(GLDv2-clean)}} &  82.28 & 75.64 & \textbf{89.34} & \textbf{79.76} & 66.57 & 51.37 & \textbf{78.56} & \textbf{61.56}\\
				\toprule
			\end{tabular}
		}	
	\end{center}
	\caption{mAP comparison against existing methods on the full benchmark. HOW:~\cite{tolias2020learning}; DSM:~\cite{Simeoni_2019_CVPR}; R-ASMK:~\cite{teichmann2019detect}; R-MAC:~\cite{tolias:hal-01842218}; GeM:~\cite{radenovic2018fine}; SOLAR:~\cite{ng2020solar}; DELG~\cite{cao2020unifying}; NetVLAD~\cite{arandjelovic2016netvlad}; +SP: spatial matching~\cite{philbin2007object}; 
		``$^\star$'': binarized local features; ``$\dagger$'': our re-implementation. 
		Training datasets are shown in brackets. PQ8 and PQ1 denote PQ quantization using 8 and 1-dimensional subspaces, respectively. R101 and R50 denote ResNet101 and ResNet50~\cite{He_2016_CVPR}.
		Underline: best previous methods. 
		Black bold: best results.}
	\label{tab:state-of-the-art}
\end{table*}

\section{Experiments} \label{sec:experiments}
\subsection{Experimental Setup}
\noindent\textbf{Training dataset}. 
The clean version of Google landmarks dataset V2 (GLDv2-clean)~\cite{weyand2020google} is used for training. 
It is first collected by Google and further cleaned by researchers from the Google Landmark Retrieval Competition 2019. 
It contains a total of 1,580,470 images and 81,313 classes. 
We randomly divide it into two subsets `train’$/$`val’ with $80\%/20\%$ split.
The `train’ split is used for training model, and the `val’ split is used for validation.

\noindent\textbf{Evaluation datasets and metrics}. 
Revisited versions of the original Oxford5k~\cite{philbin2007object} and Paris6k~\cite{philbin2008lost} datasets are used to evaluate our method, which are denoted as $\mathcal{R}$Oxf and $\mathcal{R}$Par~\cite{radenovic2018revisiting} in the following. 
Both datasets contain 70 query images and additionally include 4,993 and 6,322 database images, respectively.
Mean Average Precision (mAP) is used as our evaluation metric on both datasets with Medium and Hard protocols.
Large-scale results are further reported with the $\mathcal{R}$1M dataset, which contains one million distractor images.

\noindent\textbf{Training details}. 
All models are pre-trained on ImageNet.
For image augmentation, a $512 \times 512$-pixel crop is taken from a randomly resized image and then undergoes random color jittering.
We use a batch size of 128 to train our model on 4 NVIDIA RTX 3090 GPUs for 30 epochs, which takes about 3 days.
SGD is used to optimize the model, with an initial learning rate of 0.01, a weight decay of 0.0001, and a momentum of 0.9. 
A linearly decaying scheduler is adopted to gradually decay the learning rate to 0 when the desired number of steps is reached.
The dimension $d$ of the global feature is set as 1024.
For the ArcFace margin loss, we empirically set the margin $m$ as 0.2 and the
scale $\gamma$ as $32.0$. Refinement block number $N$ is set to 2. Test images are resized with the larger dimension equal to 1024 pixels, preserving the aspect ratio. Multiple scales are adopted, \ie, $\left[1/\sqrt{2},1,\sqrt{2} \right]$. $L_2$ normalization is applied for each scale independently, then three global features are average-pooled, followed by another $L_2$ normalization. We train each model 5 times and evaluate the one with median performance on the validation set.

\subsection{Results on Image Retrieval}
\noindent\textbf{Setting for fair comparison}.
Commonly, existing methods are compared under different settings, 
\eg, training set, backbone network, feature dimension, loss function, \etc. 
This may affect our judgment on the effectiveness of the proposed method. 
In Tab.~\ref{tab:state-of-the-art}, we re-train several methods under the same settings (using GLDv2-clean dataset and ArcFace loss, 2048 global feature dimension, ResNet101 as backbone), marked with $\dagger$. 
Based on this benchmark, we fairly compare the mAP performance of various methods and ours.

\noindent\textbf{Comparison with the state of the art}. Tab.~\ref{tab:state-of-the-art} compares our approach extensively with the state-of-the-art retrieval methods.
Our method achieves the best mAP performance in all settings.
We divide the previous methods into three groups: 

\noindent(1) \textit{Local feature aggregation}. 
The current state-of-the-art local aggregation method is R101-HOW. 
We outperform it in mAP by $1.86\%$ and $4.06\%$ on the $\mathcal{R}$Oxf dataset and by $3.91\%$, and $7.80\%$ on the $\mathcal{R}$Par dataset with Medium and Hard protocols, respectively. 
For $\mathcal{R}$1M, we also achieve the best performance.
The results show that our aggregation method is better than existing local feature aggregation methods based on large visual codebook.

\noindent(2) \textit{Global single-pass}. 
When trained with GLDv2-clean, R101-DELG achieves the best performance mostly.
When using ResNet101 as the backbone, the comparison between our method and it in mAP is 
$82.28\%$ \vs $78.24\%$, $66.57\%$ \vs $60.15\%$ on the $\mathcal{R}$Oxf dataset and
$89.34\%$ \vs $88.21\%$, $78.56\%$ \vs $76.15\%$ on the $\mathcal{R}$Par dataset with Medium and Hard protocols, respectively.
These results well demonstrate the superiority
of our framework.

\noindent(3) \textit{Global feature followed by local feature re-ranking}. 
We outperform the best two-stage method (R101-DELG+SP) in mAP by $0.50\%$, $1.80\%$ on the $\mathcal{R}$Oxf dataset and $0.88\%$, $1.76\%$ on the $\mathcal{R}$Par datasets with Medium and Hard protocols, respectively.
Although $2$-stage solutions well promote their single-stage counterparts, 
our method that aggregates local features into a compact descriptor is a better option.

\begin{figure}[t]
	\begin{center}
		\includegraphics[width=\columnwidth]{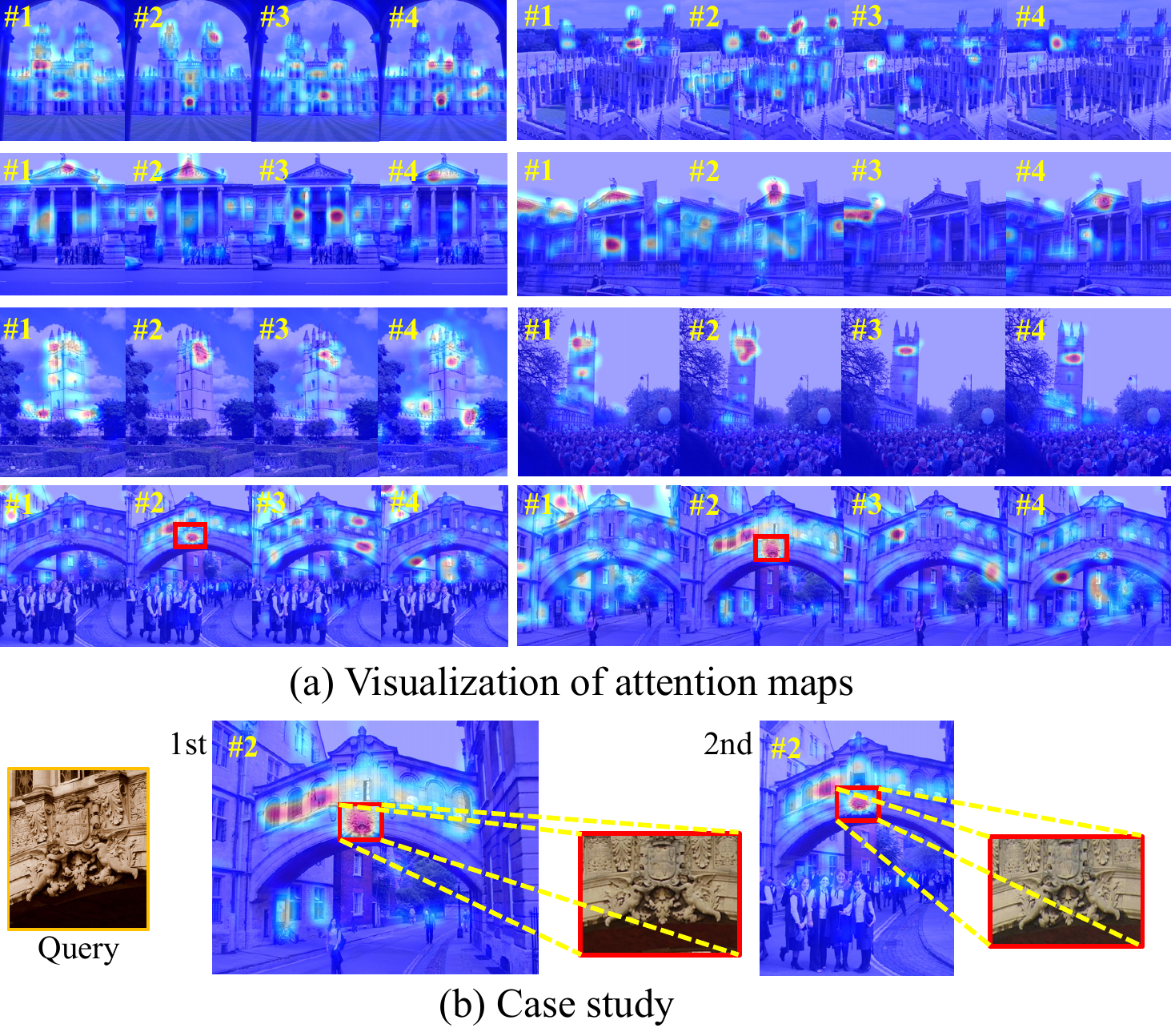}
	\end{center}
	\caption{Qualitative examples. 
		(a) Visualization of the attention maps associated with different visual tokens for eight images. $\#i$ denotes the $i$-th visual token.
		(b) Detailed analysis of the top-2 retrieval results of the ``hertford'' query in the $\mathcal{R}$Oxf dataset. The $2$nd visual token focus on the content of the query image in the target image, 
		which is boxed in red.
	}
	\label{fig:attn} 
\end{figure}

\noindent\textbf{Qualitative results}. 
To explore what the proposed tokenizer learned, we visualize the spatial attention generated by the cross-attention layer of the last refinement block in the Fig.~\ref{fig:attn}~(a).
Although there is no direct supervision, different visual tokens are associated with different visual patterns. 
Most of these patterns focus on the foreground building and remain consistent across images, which implicitly enable pattern alignment. 
\eg, the $3$rd visual token reflects the semantics of ``the upper edge of the window".

To further analyze how visual tokens improve the performance, we select the top-2 results of
the ``hertford'' query from the $\mathcal{R}$Oxf dataset for the case study. 
As shown in Fig.~\ref{fig:example}, when the query object only occupies a small part of the target image, the state-of-the-art methods with global features return false positives which are semantically similar to the query. 
Our approach uses visual tokens to distinguish different visual patterns, which has the capability of regional matching. In Fig.~\ref{fig:attn}~(b), the $2$nd visual token corresponds to the visual pattern described by the query image.

\noindent\textbf{Speed and memory costs}.
In Tab.~\ref{tab:speed_memory}, we report retrieval latency, feature extraction latency and memory footprint on $\mathcal{R}$1M for different methods.
Compared to the local feature aggregation approaches, most of the global features have a smaller memory footprint.
To perform spatial verification, ``R101-DELG+SP" needs to store a large number of local features, and thus requires about 485 GB of memory.
Our method uses a small number of visual tokens to represent the image, generating a 1024-dimensional global feature with a memory footprint of 3.9 GB. We further compress the memory requirements of global features with PQ quantization~\cite{PQ}.
As shown in Tab.~\ref{tab:state-of-the-art} and Tab.~\ref{tab:speed_memory},  the compressed features greatly reduce the memory footprint with only a small performance loss. 
Among these methods, our method appears to be a good solution in the performance-memory trade-off.

The extraction of global features is faster, since the extraction of local features usually requires scaling the image to seven scales, while global features generally use three scales.
Our aggregation method requires tokenization and iterative enhancement, which is slightly slower than direct spatial pooling, \eg, 125 ms for ours \vs 109 ms for ``R101-DELG". 
The average retrieval latency of our method on $\mathcal{R}$1M is 0.2871 seconds, which demonstrates the potential of our method for real-time image retrieval.

\begin{table}[t]
	\begin{center}
		\setlength{\tabcolsep}{3.0pt}
		\resizebox{0.473\textwidth}{!}{
			\begin{tabular}{lcccc} 
				\toprule
				\multirow{2}{*}{\Th{\ \ \ \ \ \ \ \ \large Method}} & \mr{2}{\tabincell{l}{\Th{\ \ Retrieval}\\\Th{latency (s)}}} & \mr{2}{\tabincell{l}{\Th{\ Extraction}\\\Th{latency (ms)}}} & \mc{2}{\Th{Memory} (\Th{GB})} \\ \cmidrule(l){4-5}
				&   &  & $\mathcal{R}$Oxf + $\mathcal{R}$1M & $\mathcal{R}$Par + $\mathcal{R}$1M \\ \midrule
				R50-DELF-RASMK$^\star$  & 1.5341  &   1410     & 27.6   & 27.8     \\
				R50-DELF-ASMK$^\star$   & 0.5732  &  176      & 10.3    & 10.4      \\
				R101-HOW-VLAD     & 0.4047 &      263      & 7.6    & 7.6      \\
				R101-HOW-ASMK$^\star$   &  0.7123  &  257      & 14.3   & 14.4     \\
				R101-DELG    &  0.4189  &  \textbf{109} & 7.6    & 7.6      \\
				R101-DELG+SP$^\star$  & 49.3821   &  259      & 22.6   & 22.7     \\
				R101-DELG+SP    &   58.3276   &      259      & 485.9  & 486.6    \\
				\textbf{R101-Ours}  &  0.2871  &      125      & 3.9  & 3.9      \\ 
				\textbf{R101-Ours$_{\mathbf{PQ1}}$}  & 0.2217  &      128      & 1.0  & 1.0 \\
				\textbf{R101-Ours$_{\mathbf{PQ8}}$}  & \textbf{0.1042}  &      126      & \textbf{0.1}  & \textbf{0.1}  \\ 
				\bottomrule
			\end{tabular}
		}
	\end{center}
	\caption{Time and memory measurements. We report extraction time on a
		single thread GPU (RTX 3090) / CPU (Intel Xeon CPU E5-2640 v4 @ 2.40GHz) and the search time (on a single thread CPU) for the database of $\mathcal{R}$Oxf+$\mathcal{R}$1M. }
	\label{tab:speed_memory}
\end{table}

\subsection{Ablation Study}
\noindent\textbf{Verification of different components}.
In Tab.~\ref{tad:framwork_components}, we provide experimental results to validate the contribution of the three components in our framework, 
by adding individual components to the baseline framework.
When the tokenizer is adopted, there is a significant improvement in overall performance. mAP increases from $77.0\%$ to $79.8\%$ on $\mathcal{R}$Oxf-Medium and $56.0\%$ to $62.5\%$ on $\mathcal{R}$Oxf-Hard.
This indicates that dividing local features into groups according to visual patterns is more effective than direct global spatial pooling.
From the 3rd and last row, the performance is further enhanced when the refinement block is introduced, which shows that enhancing the visual tokens with the original features further makes them more discriminative. 
There is also a performance improvement when the Local Feature Self-Attention (LFSA) is incorporated.

\begin{table}[ht]
	\begin{center}
		\small
		\setlength{\tabcolsep}{2.6pt}
		\begin{tabular}{*{7}{c}} \toprule
			\mr{2}{\Th{LFSA}} & \mr{2}{\Th{Tokenizer}} & \mr{2}{\Th{Refinement}} & \mc{2}{\Th{Medium}} & \mc{2}{\Th{Hard}} \\ \cmidrule(l){4-7}
			& & & $\mathcal{R}$Oxf & $\mathcal{R}$Par & $\mathcal{R}$Oxf & $\mathcal{R}$Par \\ \midrule
			&              &              & 77.0 & 86.6 & 56.0 & 73.0 \\
			 &   $\checkmark$  & & 79.8 & 88.2 & 62.5 & 76.0 \\
			$\checkmark$ & $\checkmark$ & & 80.4 & 88.4 & 63.0 & 76.3 \\
			
			 & $\checkmark$ & $\checkmark$ & 81.3 & 89.2 & 65.0 & 78.5 \\
			$\checkmark$  & $\checkmark$ & $\checkmark$ & \tb{82.3}  & \tb{89.3} & \tb{66.6} & \tb{78.6} \\ \bottomrule
		\end{tabular}
	\end{center}
	\caption{Ablation studies of different components. 
		We use R101-SPoC as the baseline and incrementally add tokenizer, 
		Local Feature Self-Attention (LFSA) and refinement block.}
	\label{tad:framwork_components}
\end{table}

\noindent\textbf{Impact of each component in the refinement block}.
The role of the different components in the refinement block is shown in Tab.~\ref{tab:decoder_components}. 
By removing the individual components, we find that modeling the relationship between different visual words before and further enhancing the visual tokens using the original local 
features demonstrate the effectiveness in enhancing the aggregated features.

\begin{table}[h]
	\begin{center}
		\small
		\setlength{\tabcolsep}{5.7pt}
		\begin{tabular}{*{6}{c}} \toprule
			\mr{2}{\Th{Self-Att}} & \mr{2}{\Th{Cross-Att}} & \mc{2}{\Th{Medium}} & \mc{2}{\Th{Hard}} \\ \cmidrule(l){3-6}
			& & $\mathcal{R}$Oxf & $\mathcal{R}$Par & $\mathcal{R}$Oxf & $\mathcal{R}$Par \\ \midrule
			&              & 80.4 & 88.4 & 63.0 & 76.3 \\
			& $\checkmark$ & 81.3 & 89.3 & 63.5 & 78.2 \\
			$\checkmark$  &              & 80.9 & 88.5 & 62.8 & 77.5 \\
			$\checkmark$  & $\checkmark$ & \tb{82.3}  & \tb{89.3}  & \tb{66.6} & \tb{78.6} \\ \bottomrule
		\end{tabular}
	\end{center}
	\caption{Analysis of components in the refinement block.}
	\label{tab:decoder_components}
\end{table}

\noindent\textbf{Impact of tokenizer type}.
In Tab.~\ref{tab:tokenizer}, we compare our \emph{Atten-based} tokenizer with the other two tokenizers: (1) \textit{Vq-Based}. We directly define visual tokens as a matrix $\bm{T} \in \mathcal{R}^{L \times C}$. It is randomly initialized and further updated by a moving average operation in one mini-batch. See the appendix for details.
(2) \textit{Learned}. It is similar to the \emph{Vq-Based} method, except that $\bm{T}$ is set as the network parameters, learned during training.
Our method achieves the best performance. 
We use the attention mechanism to generate visual tokens directly from the original local features.
Compared with the other two, our approach obtains more discriminative visual tokens with a better capability to match different images.
\begin{table}[t]
	\begin{center}
		\small
		\setlength{\tabcolsep}{7.0pt}
		\begin{tabular}{*{5}{c}} \toprule
			\mr{2}{\Th{Tokenizer Type}} & \mc{2}{\Th{Medium}} & \mc{2}{\Th{Hard}} \\ \cmidrule(l){2-5}
			& $\mathcal{R}$Oxf & $\mathcal{R}$Par & $\mathcal{R}$Oxf & $\mathcal{R}$Par \\ \midrule
			\Th{Vq-Based}   & 79.4 & 87.7 & 62.2 & 75.9 \\
			\Th{Learned}   & 81.1 & 87.8 & 63.7 & 76.2 \\
			\textbf{\Th{Atten-Based}}   & \tb{82.3}  & \tb{89.3}  & \tb{66.6} & \tb{78.6} \\ \bottomrule
		\end{tabular}
	\end{center}
	\caption{mAP comparison of different variants of tokenizers.}
	\label{tab:tokenizer}
\end{table}

\noindent\textbf{Impact of token number}.
The granularity of visual tokens is influenced by their number. 
As shown in Tab.~\ref{tab:tokens_number}, as $L$ increases, mAP performance first increases and then decreases, achieving the best at $L=4$. This is due to the lack of capability to distinguish local features when the number of visual tokens is small; conversely, when the number is large, they are more fine-grained and noise may be introduced when grouping local features.
\begin{table}[ht]
	\begin{center}
		\small
		\setlength{\tabcolsep}{7.0pt}
		\begin{tabular}{*{5}{c}} \toprule
			\mr{2}{\Th{Token Number}} & \mc{2}{\Th{Medium}} & \mc{2}{\Th{Hard}} \\ \cmidrule(l){2-5}
			& $\mathcal{R}$Oxf & $\mathcal{R}$Par & $\mathcal{R}$Oxf & $\mathcal{R}$Par \\ \midrule
			$L$=\Th{1}   & 79.8 & 87.9 & 60.4 & 75.7 \\
			$L$=\Th{2}   & 80.3  & 88.7  & 62.3 & 76.3 \\
			$L$=\Th{3}   & 81.6  & \tb{89.4}  & 64.9 & 78.5 \\
			$L$=\Th{4}   & \tb{82.3}  & 89.3  & \tb{66.6} & 78.6 \\
			$L$=\Th{6}   & 81.0 & 88.2 & 62.5 & \tb{78.9} \\
			$L$=\Th{8}   & 79.3 & 87.1 & 61.8 & 76.6 \\ \bottomrule
		\end{tabular}
	\end{center}
	\caption{mAP comparison of visual tokens number $L$.}
	\label{tab:tokens_number}
\end{table}

\section{Conclusion}
In this paper, we propose a joint local feature learning and aggregation framework, which generates compact global representations for images while preserving the capability of regional matching. 
It consists of a tokenizer and a refinement block. The former represents the image with a few visual tokens, which is further enhanced by the latter based on the original local features. 
By training with image-level labels, our method produces representative aggregated features. 
Extensive experiments demonstrate that the proposed method achieves superior performance on image retrieval benchmark datasets. In the future, we will extend the proposed aggregation method to a variety of existing local features, 
which means that instead of directly performing local feature learning and aggregation end-to-end, 
local features of images are first extracted using existing methods and further aggregated with our method. 

{\flushleft \bf Acknowledgements.} This work was supported in part by the National Key R$\&$D Program of China under contract 2018YFB1402605, in part by the National Natural Science Foundation of China under Contract 62102128, 61822208 and 62172381, and in part by the Youth Innovation Promotion Association CAS under Grant 2018497. It was also supported by the GPU cluster built by MCC Lab of Information Science and Technology Institution, USTC.

{
	\small
	\bibliography{aaai22}
}	
\end{document}